\documentclass[11pt]{article}
\usepackage{amssymb}
\usepackage{graphics}
\usepackage{epsfig}
\usepackage{a4wide}
\usepackage{slashed,cite}
\usepackage{color}

\begin{document}

\newcommand{\eeqq}{$e^+ e^- \to q_- \bar q_-$ }
\newcommand{\eeqqg}{$e^+ e^- \to q_- \bar q_- g$ }

\title{ Precision calculations for the $T$-odd quark pair production at the CLIC $e^+e^-$ linear collider     }
\author{  A. B. Mahfoudh, Guo Lei, Liu Wen, Ma Wen-Gan, Zhang Ren-You, and Zhang Wen-Juan\\
{\small  Department of Modern Physics, University of Science and Technology of China (USTC), }  \\
{\small   Hefei, Anhui 230026, P.R.China}}

\date{}
\maketitle \vskip 15mm
\begin{abstract}
We perform the precision calculations for the \eeqq ($q_-\bar
q_-=u_-\bar u_-, ~c_-\bar c_-,~ d_-\bar d_-,~s_-\bar s_-$) processes
up to the QCD next-to-leading order (NLO) including full weak decays
for the final $T$-odd mirror quarks in the littlest Higgs model with
$T$-parity (LHT) at the Compact Linear Collider (CLIC). We show the
dependence of the leading order (LO) and NLO QCD corrected cross
sections on the colliding energy $\sqrt{s}$, and provide the LO and QCD
NLO kinematic distributions of final particles. The results show
that the LO cross section can be enhanced by the NLO QCD correction and
the $K$-factor increases obviously when the threshold of the on-shell
$q_-\bar q_-$-pair production approaches the colliding energy $\sqrt{s}$.
The $K$-factor value varies in the range of $1.04 \sim 1.41$ in our chosen
parameter space. We find that a simple approximation of multiplying
the LO kinematic distribution with the integrated
$K$-factor is not appropriate for precision study of the \eeqq
($q_-\bar q_-=u_-\bar u_-,~c_-\bar c_-,~d_-\bar d_-,~s_-\bar s_-$)
processes, since the NLO QCD corrections are phase space dependent.
It is necessary to calculate the differential cross
sections including full NLO QCD corrections to get reliable
results.
\end{abstract}

\vskip 15mm {\bf PACS: 12.38.Bx, 13.85.Dz, 13.66.Hk  }

\vskip 5mm

{\bf Keywords: Littlest Higgs model with $T$-parity, Compact Linear Collider, NLO QCD}

\vfill \eject \baselineskip=0.32in

\renewcommand{\theequation}{\arabic{section}.\arabic{equation}}
\renewcommand{\thesection}{\Roman{section}.}
\newcommand{\nb}{\nonumber}

\newcommand{\Dir}{\kern -6.4pt\Big{/}}
\newcommand{\Dirin}{\kern -10.4pt\Big{/}\kern 4.4pt}
\newcommand{\DDir}{\kern -7.6pt\Big{/}}
\newcommand{\DGir}{\kern -6.0pt\Big{/}}

\makeatletter      
\@addtoreset{equation}{section}
\makeatother       

\par
\section{Introduction }
\par
The Compact Linear Collider (CLIC) is a high-luminosity ${\rm TeV}$ scale
$e^+e^-$ linear collider under development. It is
schemed that CLIC would provide high luminosity $e^+e^-$ collisions
from a few hundred ${\rm GeV}$ to $3~{\rm TeV}$. The first stage of
CLIC, with energy at or above $350~{\rm GeV}$, gives access to precision
Higgs physics, providing absolute values of Higgs couplings to both
fermions and bosons. This stage also addresses precision top
physics. The second stage, around $1.4~{\rm TeV}$, will open the
energy frontier allowing for the discovery of new physics phenomena.
The ultimate CLIC energy would be $3~{\rm TeV}$ which enlarges the CLIC physics
potential even further \cite{CLIC}. Therefore, this machine is
considered an important option for a post-LHC facility at CERN, as
emphasized in the recent update of the european strategy for
particle physics \cite{CLIC-1,CLIC-2}. Due to the cleaner
environment arising from $e^+ e^-$ collisions and the compelling
high energy, the CLIC could produce new heavy particles with
exciting precision, and can be expected to provide more significant
information about new physics.

\par
The Higgs boson discovery reported by the CERN Large Hadron Collider
(LHC) experiments \cite{CMS, ATLAS} indeed strengthens our
confidence in the standard model (SM)
\cite{s1,SS1,SS2,SS3,s2,SA1,SA2,SA3,SA4,SA5} which has made
remarkable success in accurately describing particle physics
including the strong and electroweak interactions. However, the
notorious hierarchy problem which describes the unstable mass of
Higgs from the SM under radiative corrections exists in the
framework of the SM. By introducing a set of new heavy gauge bosons
($W_H, Z_H, A_H$), a vector-like quark ($T$) and a Higgs triplet ($\Phi$) at the global
symmetry breaking scale $f$, the littlest Higgs (LH) model could
cancel the quadratic divergences of the Higgs mass at one-loop level
\cite{LH}. Nevertheless, the LH model suffers severe constraints from
precision electroweak measurements. The littlest Higgs model with
$T$-parity (LHT) could solve this problem successfully by bringing
in a $Z_2$ discrete symmetry described as $T$-parity, and can avoid
fine-tuning between the global symmetry breaking scale $f$ and the
electroweak symmetry breaking scale
\cite{T-parity,Low:2004xc,Hubisz:2004ft,Hubisz:2005tx}. Furthermore,
the LHT could offer a candidate for dark matter
\cite{Low:2004xc,Hubisz:2004ft,Hubisz:2005tx,Barbieri:2000gf,Cheng:2003ju}.
Thus we should pay more attention to the LHT. Recently, some QCD
next-to-leading order (NLO) phenomenological aspects of the LHT have
been analyzed in Refs.\cite{YanH, DuSM, LiuW}.

\par
In this work, we study the $q_- \bar q_-$-pair production up to the QCD NLO
in the LHT at the $e^+e^-$ CLIC including subsequential decays
of the final $T$-odd quarks. We organize this paper as follows:
In Sec.II, we describe the related LHT theory, and provide the NLO
QCD corrected decay widths of the $T$-odd quarks. In Sec.III, the
calculation strategy is presented. In Sec.IV, numerical analysis
and a discussion are provided. A short summary is given finally.

\vskip 5mm
\section{Related LHT theory }\label{theory}
\par
In the LHT scenario, 14 massless Nambu-Goldstone bosons are born out
of an $SU(5)$ global symmetry breaking down to $SO(5)$ spontaneously
at some high-scale $f$. Four of them are regarded as longitudinal
modes of the heavy gauge bosons. The other 10 decompose into a
$T$-even SU(2) doublet $h$, treated as the SM Higgs doublet, and a
complex $T$-odd $SU(2)$ triplet $\Phi$. To implement $T$-parity in
the fermion sector of the model, the mirror partners for each of the
original fermions are introduced. The masses of the $T$-odd partners of
SM up- and down-type quarks are expressed as
\begin{eqnarray} \label{m_q}
m_{u_{-}},m_{c_{-}}\simeq \sqrt{2}\kappa
f\left(1-\frac{1}{8}\frac{v^2}{f^2}\right), &&
m_{d_{-}},m_{s_{-}}=\sqrt{2}\kappa f,
\end{eqnarray}
where $\kappa$ is the mass coefficient in Lagrangian of the quark
sector, and the vacuum expectation value $v=246~{\rm GeV}$. The masses of the $T$-odd gauge bosons are given by
\begin{eqnarray}\label{mass-AH-VH}
 m_{A_H} \simeq \frac{1}{\sqrt{5}} g^{\prime} f
 \left( 1 - \frac{5}{8}\frac{v^2}{f^2} \right),~~~~
 m_{W_H} \simeq g f \left( 1 - \frac{1}{8}\frac{v^2}{f^2}
 \right),~~~~ m_{Z_H} \simeq m_{W_H}.
\end{eqnarray}

\par
The Feynman rules for the vertices in the LHT related to this work
are listed in Table \ref{tabA-1}, where $\delta_v = -\frac{v^2}{8 f^2}$,
$P_{L,R}=\frac{1}{2}(1\mp
\gamma_5)$ and $\theta_W$ is the Weinberg angle. The details of the
LHT theory can be found in
Refs.\cite{Low:2004xc,Hubisz:2004ft,Hubisz:2005tx,cpyuan:2006ph,plb670-378}.
\begin{table}[h]
\tiny
\begin{center}
\begin{tabular}{|c|l||c|l|}
\hline
Interaction & ~~~~~~~~Feynman rule & Interaction & ~~~~~~~~~Feynman rule \\
\hline
&&& \\
$\bar U_{i-} Z^\mu U_{i-}$ &
$\frac{ig}{\cos\theta_W}\left( \frac{1}{2}-\frac{2}{3}\sin^2\theta_W + \delta_v P_L \right)\gamma^\mu$
&
$\bar D_{i-} Z^\mu D_{i-}$ & $\frac{ig}{\cos\theta_W}\left( -\frac{1}{2}+\frac{1}{3}\sin^2\theta_W \right)\gamma^\mu$\\
&&& \\
$\bar U_{i-} A^\mu U_{i-}$ & $ieQ_{U_{i-}}\gamma^\mu$ &
$\bar D_{i-} A^\mu D_{i-}$ & $ieQ_{D_{i-}}\gamma^\mu$ \\
&&& \\
$\bar{U}_{i-} A_{H}^{\mu} U_j$ & $i\left( -\frac{g'}{10}\cos\theta_H-\frac{g}{2}
\sin\theta_H \right)(V_{Hu})_{ij} \gamma^\mu P_L$ &
$\bar{D}_{i-} A_{H}^{\mu} D_j$ & $i\left( -\frac{g'}{10}\cos\theta_H+\frac{g}{2}
\sin\theta_H \right)(V_{Hd})_{ij} \gamma^\mu P_L$\\
&&&\\
$\bar{U}_{i-}Z_{H}^{\mu} U_j$ & $i\left( -\frac{g'}{10}\sin\theta_H+\frac{g}{2}\cos\theta_H
\right)(V_{Hu})_{ij} \gamma^\mu P_L$ &
$\bar{D}_{i-}Z_{H}^{\mu} D_j$ & $i\left( -\frac{g'}{10}\sin\theta_H-\frac{g}{2}\cos\theta_H
\right)(V_{Hd})_{ij} \gamma^\mu P_L$\\
&&& \\
$\bar D_{i-} W^{-\mu}_{H} U_{j}$ & $i\frac{g}{\sqrt{2}} (V_{Hu})_{ij}\gamma^\mu P_L$ &
$\bar U_{i-} W^{+\mu}_{H} D_{j}$ & $i\frac{g}{\sqrt{2}}
(V_{Hd})_{ij}\gamma^\mu P_L$ \\
&&&\\
$\bar{q}_{-}^{\alpha} q_{-}^{\beta} G^{a}_{\mu}$ & $ig_s (T^a)_{\alpha\beta}\gamma^{\mu}$ && \\
&&& \\
\hline
\end{tabular}
\caption{\label{tabA-1} The related LHT Feynman rules used in this
work, where $U_{i-}=u_-,c_-$ and $D_{i-}=d_-,s_-$, $i,j = 1,2$ are the generation
indices. }
\end{center}
\end{table}

\par
The LO partial decay widths of $T$-odd up- and down-type mirror quarks can
be found in Appendix B of Ref.\cite{DuSM}. We calculate the NLO QCD
corrected partial decay widths of $T$-odd up- and down-type mirror quarks
of the first two generations in case of neglecting light quark
masses and the terms of order $(\alpha_s/\pi)m_{V_H}^2/m_{q_-}^2$
$(V_H=Z_H,W_H,A_H)$, which can be accepted in the parameter space
adopted in this work. The explicit expressions for the relevant
partial decay widths at NLO can be written as
\begin{eqnarray} \label{Width-1}
\Gamma (U_{i-}\to A_H U_j) &=& \frac{2|(V_{Hu})_{ij}|^2\left(\frac{g
s_H}{2} + \frac{g' c_H}{10}\right)^2}{64\pi}\frac{m_{U_{i-}}^3}
{m_{A_H}^2} \left(1- \frac{m_{A_H}^2}{m_{U_{i-}}^2}\right)^2
\left(1+ \frac{2m_{A_H}^2}{m_{U_{i-}}^2}\right)  \nb\\
&&  \left[1-\frac{2\alpha_s}{3\pi}
\left(\frac{2\pi^2}{3}-\frac{5}{2}\right)\right],~~~~~~~~~~~~~(i,j=1,2), \nb\\
\Gamma (U_{i-}\to Z_H U_j) &=& \frac{2|(V_{Hu})_{ij}|^2\left(\frac{g
c_H}{2} - \frac{g' s_H}{10}\right)^2}{64\pi}\frac{m_{U_{i-}}^3}
{m_{Z_H}^2} \left(1- \frac{m_{Z_H}^2}{m_{U_{i-}}^2}\right)^2
\left(1+ \frac{2m_{Z_H}^2}{m_{U_{i-}}^2}\right)  \nb\\
&&  \left[1-\frac{2\alpha_s}{3\pi}
\left(\frac{2\pi^2}{3}-\frac{5}{2}\right)\right], ~~~~~~~~~~~~~(i,j=1,2), \nb\\
\Gamma (U_{i-}\to W_H^+ D_j) &=& \frac{g^2|(V_{Hd})_{ij}|^2}{64\pi}\frac{m_{U_{i-}}^3}
{m_{W_H}^2} \left(1- \frac{m_{W_H}^2}{m_{U_{i-}}^2}\right)^2
\left(1+ \frac{2m_{W_H}^2}{m_{U_{i-}}^2}\right)  \nb\\
&&  \left[1-\frac{2\alpha_s}{3\pi}
\left(\frac{2\pi^2}{3}-\frac{5}{2}\right)\right], ~~~~~~~~~~~~~(i,j=1,2), \nb\\
\Gamma (D_{i-}\to A_H D_j) &=& \frac{2|(V_{Hd})_{ij}|^2\left(\frac{g
s_H}{2} - \frac{g' c_H}{10}\right)^2}{64\pi}\frac{m_{D_{i-}}^3}
{m_{A_H}^2}\left(1- \frac{m_{A_H}^2}{m_{D_{i-}}^2}\right)^2 \left(1+
\frac{2m_{A_H}^2}{m_{D_{i-}}^2}\right)   \nb\\
&&  \left[1-\frac{2\alpha_s}{3\pi}
\left(\frac{2\pi^2}{3}-\frac{5}{2}\right)\right]£¬ ~~~~~~~~~~~~~(i,j=1,2), \nb\\
\Gamma (D_{i-}\to Z_H D_j) &=& \frac{2|(V_{Hd})_{ij}|^2\left(\frac{g
c_H}{2} + \frac{g' s_H}{10}\right)^2}{64\pi}\frac{m_{D_{i-}}^3}
{m_{Z_H}^2}\left(1- \frac{m_{Z_H}^2}{m_{D_{i-}}^2}\right)^2 \left(1+
\frac{2m_{Z_H}^2}{m_{D_{i-}}^2}\right)   \nb\\
&&  \left[1-\frac{2\alpha_s}{3\pi}
\left(\frac{2\pi^2}{3}-\frac{5}{2}\right)\right]£¬~~~~~~~~~~~~~(i,j=1,2), \nb\\
\Gamma (D_{i-}\to W_H^- U_j) &=& \frac{g^2|(V_{Hu})_{ij}|^2}
{64\pi}\frac{m_{D_{i-}}^3}
{m_{W_H}^2}\left(1- \frac{m_{W_H}^2}{m_{D_{i-}}^2}\right)^2 \left(1+
\frac{2m_{W_H}^2}{m_{D_{i-}}^2}\right)  \nb\\
&&  \left[1-\frac{2\alpha_s}{3\pi}
\left(\frac{2\pi^2}{3}-\frac{5}{2}\right)\right]£¬ ~~~~~~~~~~~~~(i,j=1,2),
\end{eqnarray}
where $U_{i-}=u_-,c_-$, $D_{i-}=d_-,s_-$, $U_i=u,c$, $D_i=d,s$,
$s_{H}=\sin{\theta_{H}}$, $c_{H}=\cos{\theta_{H}}$, and the mixing
angle $\theta_H$ at the ${\cal O}(v^2/f^2)$ is expressed as
\begin{eqnarray}
\sin \theta_H \simeq \left[ \frac{5gg'}{4(5g^2-g'^2)}\frac{v
^2}{f^2} \right].
\end{eqnarray}
As discussed in Ref.\cite{Blanke:2007ckm}, the two mixing matrices
satisfy $V_{Hu}^{\dag}V_{Hd}=V_{CKM}$. Therefore, they cannot
simultaneously be set to the identity. In the following calculations
we take $V_{Hu}$ to be a unit matrix, then we have $V_{Hd}=V_{CKM}$.
The NLO QCD corrected total decay width of the $T$-odd quark $q_-$ can be
obtained approximately by summing up all the NLO partial decay widths of the
main decay channels shown in Eqs.(\ref{Width-1}).

\vskip 5mm
\par
\section{Calculations }\label{calc}
\par
We employ the FeynArts 3.4 package \cite{FeynArts} to generate
Feynman diagrams and their corresponding amplitudes in the LO and
QCD NLO calculations. The t'Hooft-Feynman gauge is adopted in this
work except when we verify the gauge invariance. The reduction of
the amplitudes are implemented by the FormCalc 5.4 programs
\cite{FormCalc}.

\par
\subsection{LO cross section }
\par
\begin{figure}
\begin{center}
\includegraphics[width=0.4\textwidth]{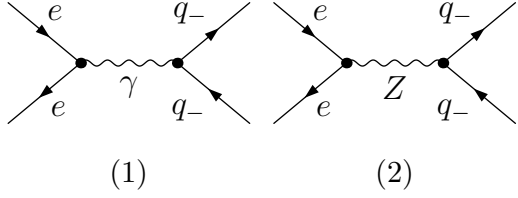}
\caption{ \label{fig1} The lowest order Feynman diagrams for
the \eeqq process in the LHT. }
\end{center}
\end{figure}

The contribution to the cross section of process \eeqq in the
LHT at the lowest order is of $\cal O$($\alpha_{ew}^2$)
with pure electroweak interactions.
We present the tree-level Feynman diagrams in Fig.\ref{fig1}.
The \eeqq process at the CLIC can be denoted as
\begin{eqnarray}
\label{process} e^{+}(p_{1})+e^{-}(p_{2})\to
q_-(p_{3})+\bar q_-(p_{4}), && (q_- \bar q_-= u_- \bar u_-,
c_- \bar c_-, d_- \bar d_-, s_- \bar s_-)
\end{eqnarray}
where $p_i~(i = 1, 2, 3, 4)$ represent the four-momenta of the
incoming and outgoing particles. The differential cross section
for the process \eeqq at the tree-level with unpolarized incoming
particles can be obtained as
\begin{eqnarray}
d{\sigma}^0_{q_- \bar q_-}= \frac{1}{4}\frac{(2\pi)^4N_{c}}{4 |\vec{p}_{1}| \sqrt{s}}
\sum_{spin} |{\cal M}^{0}_{q_- \bar q_-}|^2 d\Phi_2,
\end{eqnarray}
where the color number $N_c=3$ and the summation is taken over the
spins of initial and final particles. The factor $\frac{1}{4}$ is
due to taking average over the polarization states of the electron
and positron. $d\Phi_2$ is the two particle phase space element
defined as
\begin{eqnarray}
d\Phi_2=\delta^{(4)} \left( p_1+p_2-p_3-p_4 \right) \prod_{i=3}^4
\frac{d^3 \vec{p}_i}{(2 \pi)^3 2 E_i}.
\end{eqnarray}
In this work we provide the total cross section and distributions
for all the $(u_- \bar u_-), (c_- \bar c_-), (d_- \bar d_-), (s_- \bar s_-)$
pair production processes. The LO total cross section for processes (\ref{process})
can be figured out from the following formula:
\begin{eqnarray}
\sigma_{LO} &=& \sum_{q_-=u_-,d_-,c_-,s_-}
\sigma^{0}_{q_- \bar q_-}.
\end{eqnarray}

\par
\subsection{QCD NLO corrections to \eeqq process }
\par
\begin{figure}
\begin{center}
\includegraphics[width=0.4\textwidth]{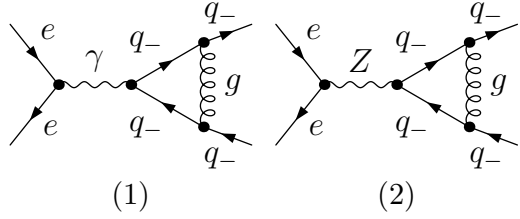}
\caption{ \label{fig2} The representative one-loop Feynman diagrams
for the process \eeqq. }
\end{center}
\end{figure}

The QCD NLO corrections to the process \eeqq at the CLIC can be
divided into two parts: (i) The QCD one-loop virtual corrections to
the processes \eeqq; (ii) The contributions of the real gluon
emission process \eeqqg. In the NLO calculations we adopt the
dimensional regularization scheme, in which the dimensions of spinor
and space-time manifolds are extended to $D=4-2\epsilon$ to isolate
the ultraviolet (UV) and infrared (IR) singularities.
The representative Feynman diagrams for the one-loop virtual
corrections to the process \eeqq are presented in Fig.\ref{fig2}.
There exist both UV and IR singularities. The masses and wave
functions of $T$-odd partners of SM quarks should be renormalized to
remove the UV divergences. The counterterms are defined as
\begin{eqnarray}
\psi^{0,L,R}_{q_-} & = & \left(1+\frac{1}{2}
\delta Z_{q_-}^{L,R}\right)\psi^{L,R}_{q_-}~, \nb \\
m^0_{q_-} & = & m_{q_-}+\delta m_{q_-}~,
\end{eqnarray}
where $\psi^{L,R}_{q_-}$ denote the fields of $T$-odd mirror quark, and
$m_{q_-}$denotes the mass of $T$-odd mirror quark. Taking the on-mass-shell
renormalized condition we get the renormalization constants as
\begin{eqnarray}
\delta Z^{L,R}_{q_-} & = & - \frac{\alpha_s (\mu_r)}{3\pi}
\left[\Delta_{UV} + 2\Delta_{IR} + 4 + 3\ln
\left(\frac{\mu_r^2}{m_{q_-}^2} \right) \right]~ , \nb \\
\frac{\delta m_{q_-}}{m_{q_-}} & = & - \frac{\alpha_s
(\mu_r)}{3\pi}\left\{3\left[\Delta_{UV}+
\ln\left(\frac{\mu_r^2}{m_{q_-}^2}\right)\right]+4\right\}~.
\end{eqnarray}
There $\Delta_{UV}=\frac{1}{\epsilon_{UV}}-\gamma_E + \ln (4\pi)$
and $\Delta_{IR}=\frac{1}{\epsilon_{IR}}-\gamma_E + \ln (4\pi)$. The
one-loop virtual contribution is UV finite after performing the
renormalization procedure. The remaining IR divergencies can be
cancelled by the real gluon bremsstrahlung corrections in the soft
gluon limit, as we shall see later.

\par
\begin{figure}
\begin{center}
\includegraphics[width=0.7\textwidth]{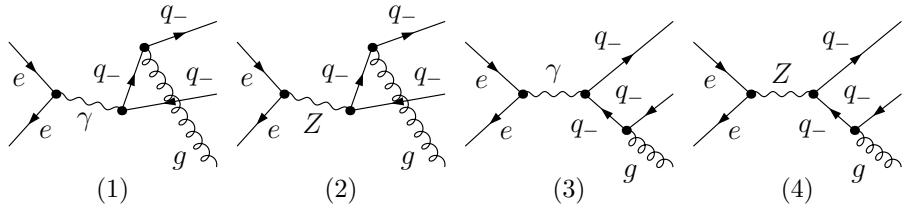}
\caption{ \label{fig3} The tree-level Feynman diagrams
for the real gluon emission process \eeqqg. }
\end{center}
\end{figure}

The real gluon emission process is denoted as
\begin{eqnarray}
\label{real g emission}
 e^+(p_1)+e^-(p_2) \to q_-(p_3)+ \bar q_-(p_{4})+g(p_5), &&
(q_- \bar q_-= u_- \bar u_-,c_- \bar c_-,d_- \bar d_-,s_- \bar s_-),
\end{eqnarray}
where the real gluon radiates from $q_-$($\bar q_{-}$) line. We
adopt the two-cutoff phase space slicing (TCPSS) method\cite{tcpss}
to isolate the IR soft singularity. The tree-level Feynman diagrams
for this process are shown in Fig.\ref{fig3}. In performing the
calculations in this work with the TCPSS method, an arbitrary small
soft cutoff $\delta_s$ should be introduced. The phase space of the
\eeqqg process can be split into two regions: soft gluon region
($E_5\leq \frac{1}{2}\delta_s \sqrt{s}$) and hard gluon region
($E_5> \frac{1}{2}\delta_s \sqrt{s}$) by the soft cutoff
$\delta_s$. Thus the cross section for the real gluon emission
process is decomposed into soft and hard noncollinear terms, i.e.,
$\sigma^R_g =\sigma^S_g+\sigma^H_g$.
According to the Kinoshita-Lee-Nauenberg (KLN) theorem \cite{KLN},
the soft singularity in the soft part $\sigma_g^{S}$ can be
canceled by the soft IR divergence in the virtual corrections while
the hard cross section part $\sigma_g^{H}$ is IR safe.

\par
After eliminating all the UV and IR singularities by performing the
renormalization procedure and adding all the QCD NLO correction
components, we can get the finite QCD NLO corrected integrated cross
section for the \eeqq process as
\begin{eqnarray}
\label{TotalCorr}
\sigma_{NLO}&=&\sigma_{LO}+\Delta
\sigma_{NLO}= \sigma_{LO}+\Delta\sigma^{(2)}+\Delta\sigma^{(3)}.
\end{eqnarray}
The two-body term $\Delta \sigma^{(2)}$ includes the one-loop
corrections to the \eeqq process and the tree-level contributions
in the soft region for the real gluon emission processes,
while the three-body term $\Delta \sigma^{(3)}$ contains the
cross sections for the real gluon emission processes over the
hard noncollinear region.

\vskip 5mm
\section{Numerical results and discussions}
\par
\subsection{Input parameters}\label{parameters}
\par
The global symmetry breaking scale $f$ is an important parameter of
the LHT and has been constrained by recent experiments. Combining
all direct LHT new particle searches, an exclusion at $95\%$ CL on
the scale $f$ of $f \lesssim 620~{\rm GeV}$ is presented \cite{Snowmass}.
In our calculations we take the value of $f$ being around this
limitation. The CKM matrix is set to be the unit matrix. The other
relevant input parameters are chosen as \cite{databook}
$\alpha_{{\rm ew}}^{-1}=137.036$, $m_W=80.385~{\rm GeV}$,
$m_Z=91.1876~{\rm GeV}$, $m_e=0.511~{\rm MeV}$. We define
$\mu_0=m_{q_-}$ and fix the $T$-odd mirror quark mass coefficient $\kappa$
to be $1$ in case no other statement. As a result, the masses of
$T$-odd mirror quarks are only the functions of the LHT parameter $f$ as
shown in Eq.(\ref{m_q}). By using Eq.(\ref{m_q}) and taking the LHT
parameter $\kappa = 1$, we obtain the masses of $T$-odd mirror quarks for
some typical values of the LHT global symmetry breaking scale $f$
listed in Table \ref{tab1}.
\begin{table}
\begin{center}
\begin{tabular}{c|c|c|c|c}
  \hline
    $f$    & $m_{u_-}=m_{c_-}$  & $m_{d_-}=
m_{s_-}$  & $m_{W_H}\approx m_{Z_H}$  & $m_{A_H}$ \\
   $~~({\rm GeV})~~$ & $~~({\rm GeV})~~$ & $~~({\rm GeV})~~$
   &  $~~({\rm GeV})~~$  &  $~~({\rm GeV})~~$\\
  \hline
     600  & 830.7   & 848.5    & 376.8  &   82.5 \\
     700  & 974.7   & 989.9    & 442.1  &   99.2 \\
     800  & 1118.0  & 1131.4   & 507.1  &  115.6 \\
     900  & 1260.9  & 1272.8   & 571.9  &  131.8 \\
     1000 & 1403.5  & 1414.2   & 636.6  &  147.8 \\
     1100 & 1545.9  & 1555.6   & 701.1  &  163.7 \\
     1200 & 1688.1  & 1697.1   & 765.6  &  179.5 \\
     1300 & 1830.3  & 1838.5   & 830.1  &  195.2 \\
     1400 & 1972.3  & 1979.9   & 894.5  &  210.9 \\
     1500 & 2114.2  & 2121.3   & 958.9  &  226.6 \\
     1600 & 2256.1  & 2262.7   &1023.2  &  242.2 \\
  \hline
\end{tabular}
\end{center}
\begin{center}
\begin{minipage}{15cm}
\caption{\label{tab1} The masses of $T$-odd mirror quarks $q_-$
($q_-=u_-,d_-,c_-,s_-$) and heavy vector boson ($W_H,~Z_H,~A_H$) for some typical values of
 the LHT parameter $f$ taking $\kappa =1$. }
\end{minipage}
\end{center}
\end{table}

\par
\subsection{Checks}
\par
The correctness of our calculations are verified through the
following aspects:

\par
{\bf 1.} After combining all the contributions at the QCD NLO,
the cancelations of UV and IR divergences are verified numerically.

\par
\begin{figure}
\begin{center}
\includegraphics[scale=0.73]{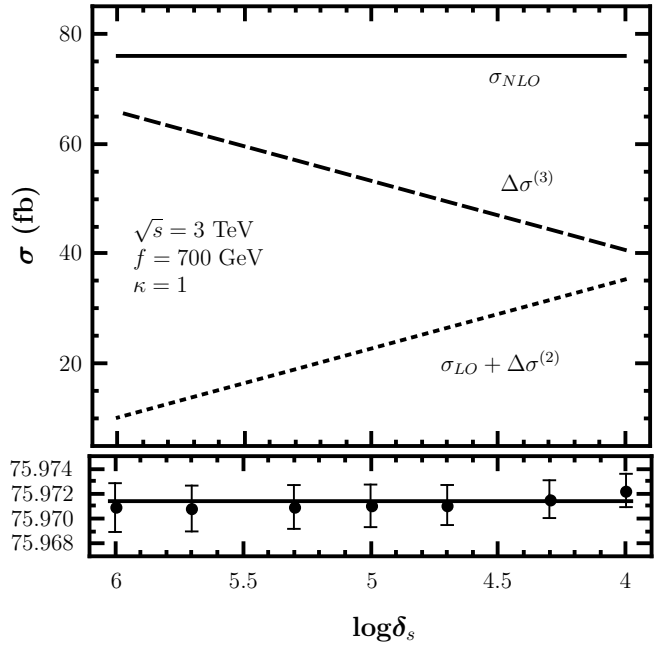}
\caption{ \label{fig4} The dependence of the NLO QCD corrected cross section
for \eeqq ($q_-\bar q_-=u_-\bar u_-,~c_-\bar c_-,~d_-\bar d_-,~s_-\bar s_-$)
on the cutoff $\delta_s$ at the
$\sqrt{s}=3~{\rm TeV}$ CLIC, and the amplified curve for
$\sigma_{tot}(=\sigma_{NLO})$ with the calculation errors are shown in lower plot.}
\end{center}
\end{figure}

{\bf 2.} We make the verification of the independence
of the total NLO QCD correction on $\delta_s$, where an arbitrary cutoff
$\delta_s$ is introduced to separate the phase space in order to isolate
the soft IR divergences \cite{tcpss}. Eq.(\ref{TotalCorr}) shows that
the total NLO QCD corrected cross section ($\sigma_{NLO}$) for all the four processes
\eeqq ($q_-\bar q_-=u_-\bar u_-,~c_-\bar c_-,~d_-\bar d_-,~s_-\bar s_-$) is obtained by
summing up the two-body and three-body contribution parts
($\sigma_{LO}+\Delta \sigma^{(2)}$ and $\Delta \sigma^{(3)}$). We depict
$\sigma_{LO}+\Delta \sigma^{(2)}$, $\Delta \sigma^{(3)}$ and $\sigma_{tot}
(=\sigma_{NLO})$ for the \eeqq ($q_-\bar q_-=u_-\bar u_-,~
c_-\bar c_-,~d_-\bar d_-,~s_-\bar s_-$) processes at the $\sqrt{s}=3~{\rm TeV}$ CLIC
as functions of the soft cutoff $\delta_s$ in the upper part of Fig.\ref{fig4}
with $f=700~{\rm GeV}$, $\kappa=1$ and renormalization scale $\mu=\mu_0=m_{q_-}
=974.7,~974.7,~989.9,~989.9~{\rm GeV}$ for $u_-\bar u_-$-, $c_-\bar c_-$-,
$d_-\bar d_-$- and $s_-\bar s_-$-pair production processes, respectively.
The amplified curve for the total NLO QCD corrected cross section $\sigma_{tot}$
in the upper plot of Fig.\ref{fig4}, is shown in the lower
plot of Fig.\ref{fig4} together with calculation errors.
From the figure we can see that the total NLO QCD corrected cross section
$\sigma_{NLO}$ is independent of the cutoff within the
statistical errors. That is an indirect check for the
correctness of our calculation. In further numerical calculations,
we set $\delta_s = 1\times 10^{-4}$ .

\par
\subsection{Dependence on colliding energy $\sqrt{s}$}
\par
\begin{figure}
\begin{center}
\includegraphics[scale=0.74]{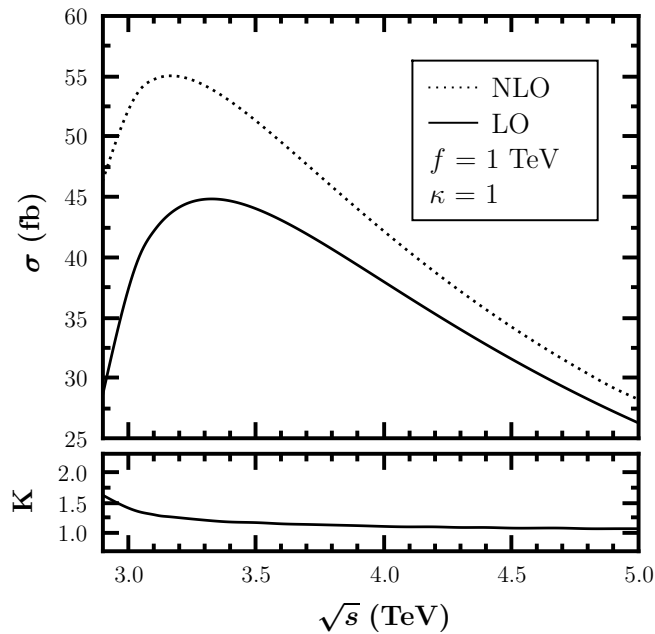}
\caption{ \label{fig5} The LO and NLO QCD corrected cross sections
for \eeqq ($q_-\bar q_-=u_-\bar u_-,~c_-\bar c_-,~d_-\bar d_-,~s_-\bar s_-$)
as functions of colliding energy
$\sqrt{s}$ with $f=1000~{\rm GeV}$ and $\kappa = 1$.
The lower plot shows the corresponding $K$-factor
($K \equiv \frac{\sigma_{NLO}}{\sigma_{LO}}$) as the
function of $\sqrt{s}$. }
\end{center}
\end{figure}

In the upper part of Fig.\ref{fig5} we present the LO and ${\cal
O}(\alpha_{s})$ QCD corrected cross sections ($\sigma_{LO}$,
$\sigma_{NLO}$) for the \eeqq ($q_-\bar q_-=u_-\bar u_-,~c_-\bar
c_-,~d_-\bar d_-,~s_-\bar s_-$) processes as the functions of
colliding energy $\sqrt{s}$ with $f=1000~{\rm GeV}$ and $\kappa = 1$. The
corresponding $K$-factor defined as $K\equiv
\frac{\sigma_{NLO}}{\sigma_{LO}}$, is presented in the lower plot of
Fig.\ref{fig5}. We can see from Fig.\ref{fig5} that both the LO and
NLO QCD corrected cross sections are sensitive to the colliding
energy, particularly in the range of $\sqrt{s}\in [2900,~3300]~{\rm GeV}$ due
to the threshold effect, while decrease quickly when
$\sqrt{s}>3300~{\rm GeV}$. In the lower plot of Fig.\ref{fig5} it shows
that the $K$-factor has a large value in the vicinity where the
colliding energy is close to the $q_-\bar q_-$ threshold due to
Coulomb singularity effect.

\par
\subsection{Dependence on global symmetry breaking scale $f$}
\par
\begin{figure}[htbp]
\begin{center}

\includegraphics[scale=0.64]{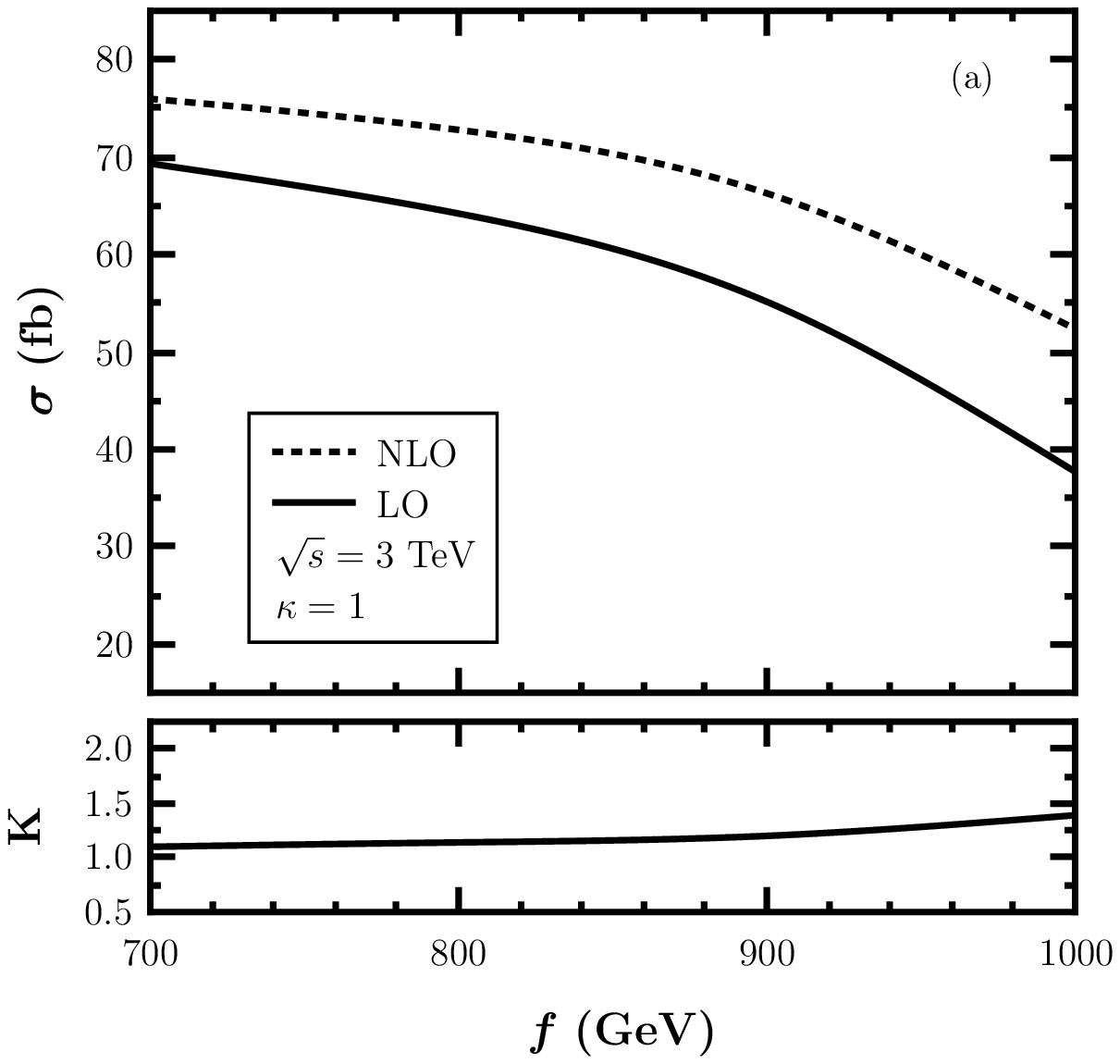}
\includegraphics[scale=0.64]{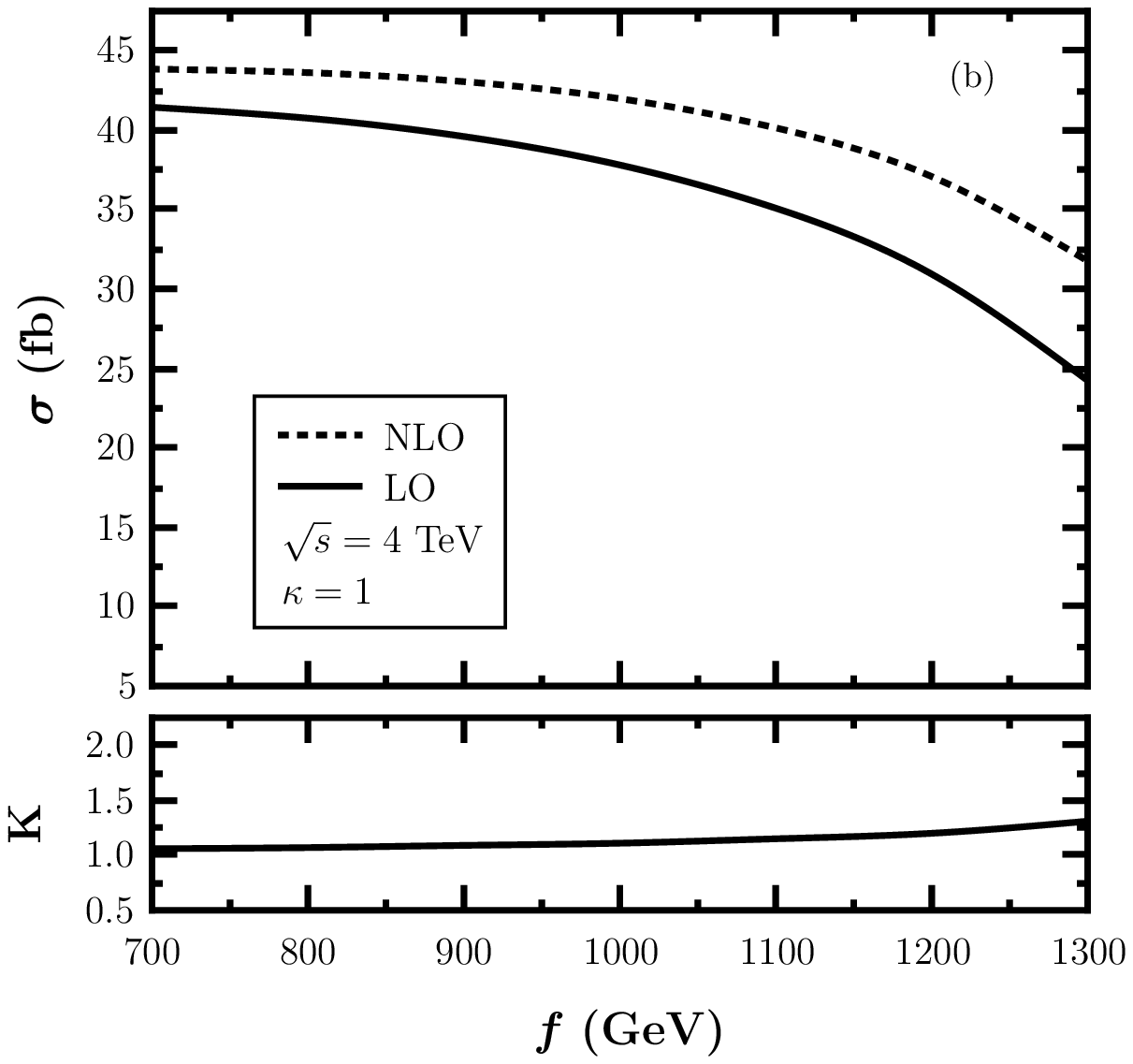}
\\~~  \\
\includegraphics[scale=0.64]{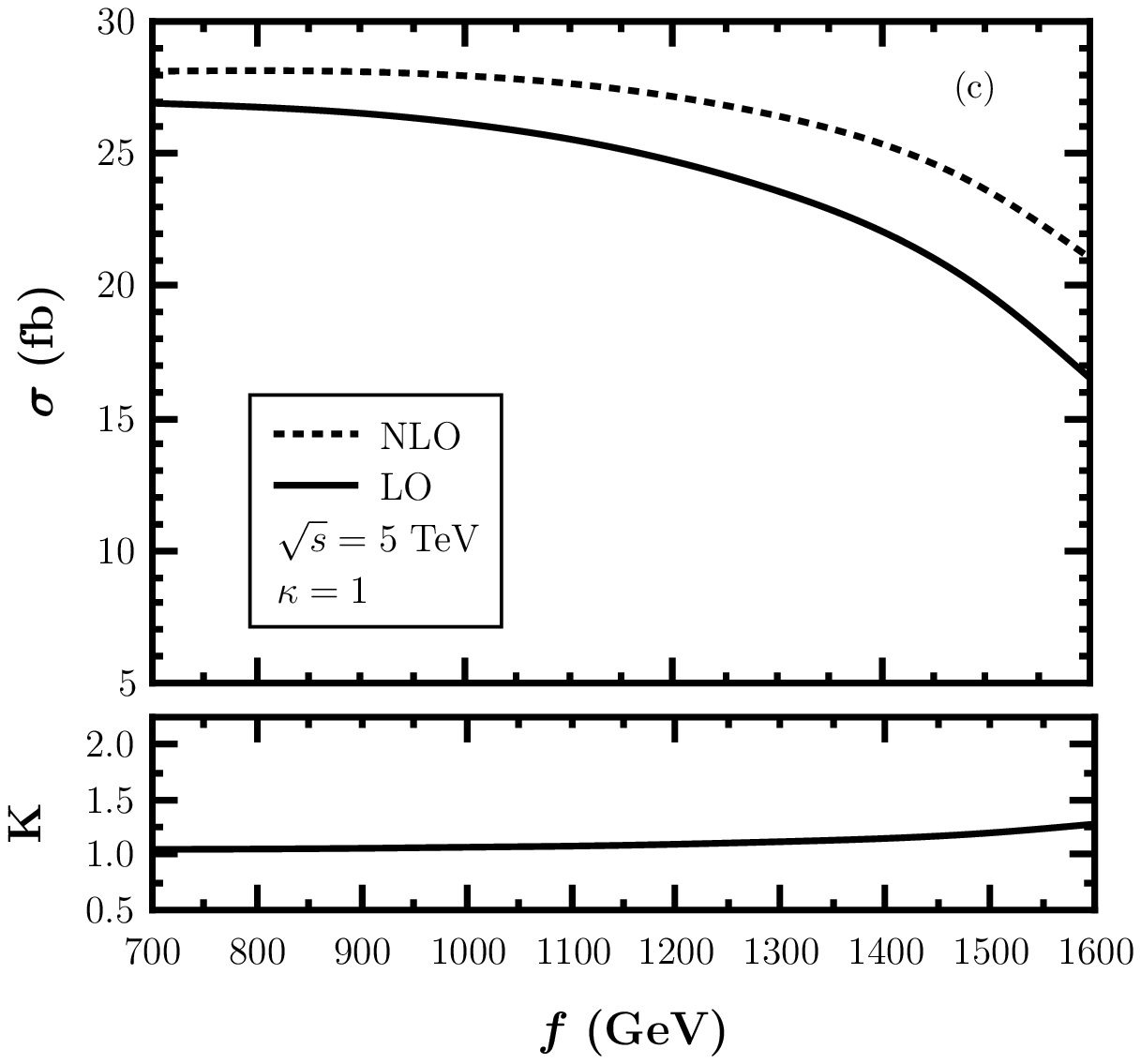}
\hspace{0in}%
\caption{\label{fig6} The LO, NLO QCD corrected cross sections and
the corresponding $K$-factor for \eeqq ($q_-\bar q_-=u_-\bar
u_-,~c_-\bar c_-,~d_-\bar d_-,~s_-\bar s_-$) as
functions of the global symmetry breaking scale $f$ with $\kappa
=1$. (a) $\sqrt{s}=3~{\rm TeV}$ CLIC. (b) $\sqrt{s}=4~{\rm TeV}$
CLIC. (c) $\sqrt{s}=5~{\rm TeV}$ CLIC. }
\end{center}
\end{figure}

We depict the LO, NLO QCD corrected integrated cross sections and
the corresponding $K$-factors as functions of the global
symmetry breaking scale $f$ by taking $\kappa=1$ at the
$\sqrt{s}=3~{\rm TeV}$, $4~{\rm TeV}$ and $5~{\rm TeV}$ CLIC in Figs.\ref{fig6}(a),
(b) and (c), respectively. Figs.\ref{fig6}(a,b,c) demonstrate that
the LO and NLO QCD corrected total cross sections for the \eeqq
($q_-\bar q_-=u_-\bar u_-,~c_-\bar c_-,~d_-\bar d_-,~s_-\bar s_-$)
processes decrease with the increment of $f$ due to the fact that
the masses of final $q_-$ and $\bar q_-$ become heavier and
consequently the phase space becomes smaller as the increment of
$f$. We see that the curves for $K$-factors in the lower plots of
Figs.\ref{fig6}(a,b,c) increase slightly with the increament of scale
$f$ from $700~{\rm GeV}$ to $1000~{\rm GeV}$. The reason is that the
threshold value ($2 m_{q_-}$) approaches to the colliding energy
with the increment of scale $f$. The numerical results for the \eeqq
($q_-\bar q_-=u_-\bar u_-,~ c_-\bar c_-,~d_-\bar d_-,~s_-\bar s_-$)
processes at the CLIC for some typical values of $f$ are presented
in Table \ref{tab2}.
\begin{table}
\begin{center}
\begin{tabular}{c|c|c|c|c}
\hline
$\sqrt{s}$ ({\rm TeV}) & $f$ ({\rm TeV}) &
$\sigma_{LO}$ ({\rm fb}) & $\sigma_{NLO}$ ({\rm fb})&$K$ \\
\hline
    & 700   & 69.32658(3)   & 75.9723(5)  & 1.10      \\
  3 & 800   & 64.19684(3)  & 72.7819(2) & 1.14      \\
    & 900   & 55.13894(1)  & 66.2727(1) & 1.20      \\
    & 1000  & 37.688101(6) & 52.36895(1) & 1.39      \\
\hline
    & 700   & 41.44320(2)  & 43.8441(4)  & 1.06      \\
  4 & 1000  & 37.80775(1) & 41.9819(2) & 1.11      \\
    & 1200  & 30.93860(5) & 37.08019(1)& 1.20      \\
    & 1300  & 24.241070(4) & 31.76587(6) & 1.31      \\
\hline
    & 700   & 26.90824(7)  & 28.1171(2) & 1.05      \\
    & 900   & 26.52824(9)  & 28.0985(3) & 1.06      \\
    & 1000  & 26.13502(8)  & 27.9454(1) & 1.07      \\
  5 & 1100  & 25.55674(5)  & 27.65382(8) & 1.08      \\
    & 1300  & 23.59325(6)  & 26.42949(9) & 1.12      \\
    & 1500  & 19.77699(7) & 23.64070(9) & 1.20      \\
    & 1600  & 16.52192(3) & 21.06044(4) & 1.28      \\
\hline
\end{tabular}
\end{center}
\begin{center}
\begin{minipage}{15cm}
\caption{\label{tab2} The numerical results of $\sigma_{LO}$,
$\sigma_{NLO}$ and the corresponding $K$-factors for the \eeqq
($q_-\bar q_-=u_-\bar u_-,~c_-\bar c_-,~d_-\bar d_-,~s_-\bar s_-$) processes
with different values of $f$ and colliding energy
$\sqrt{s}$ at the CLIC by taking $\kappa = 1$ and $\mu=\mu_0$.  }
\end{minipage}
\end{center}
\end{table}

\par
\subsection{Dependence on $T$-odd mirror quark mass coefficient $\kappa$}
\par
\begin{figure}[htbp]
\begin{center}
\includegraphics[scale=0.64]{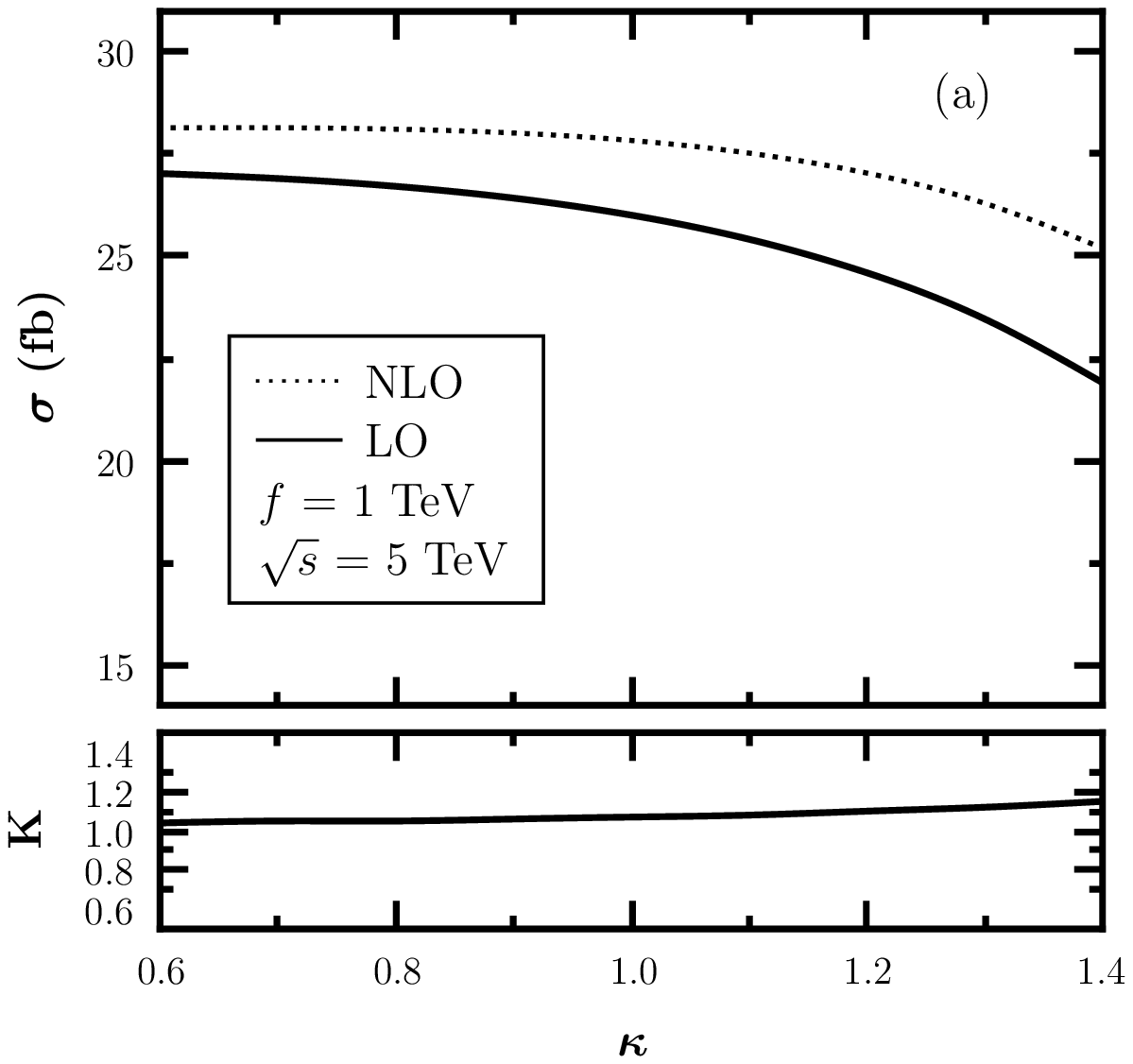}
\includegraphics[scale=0.64]{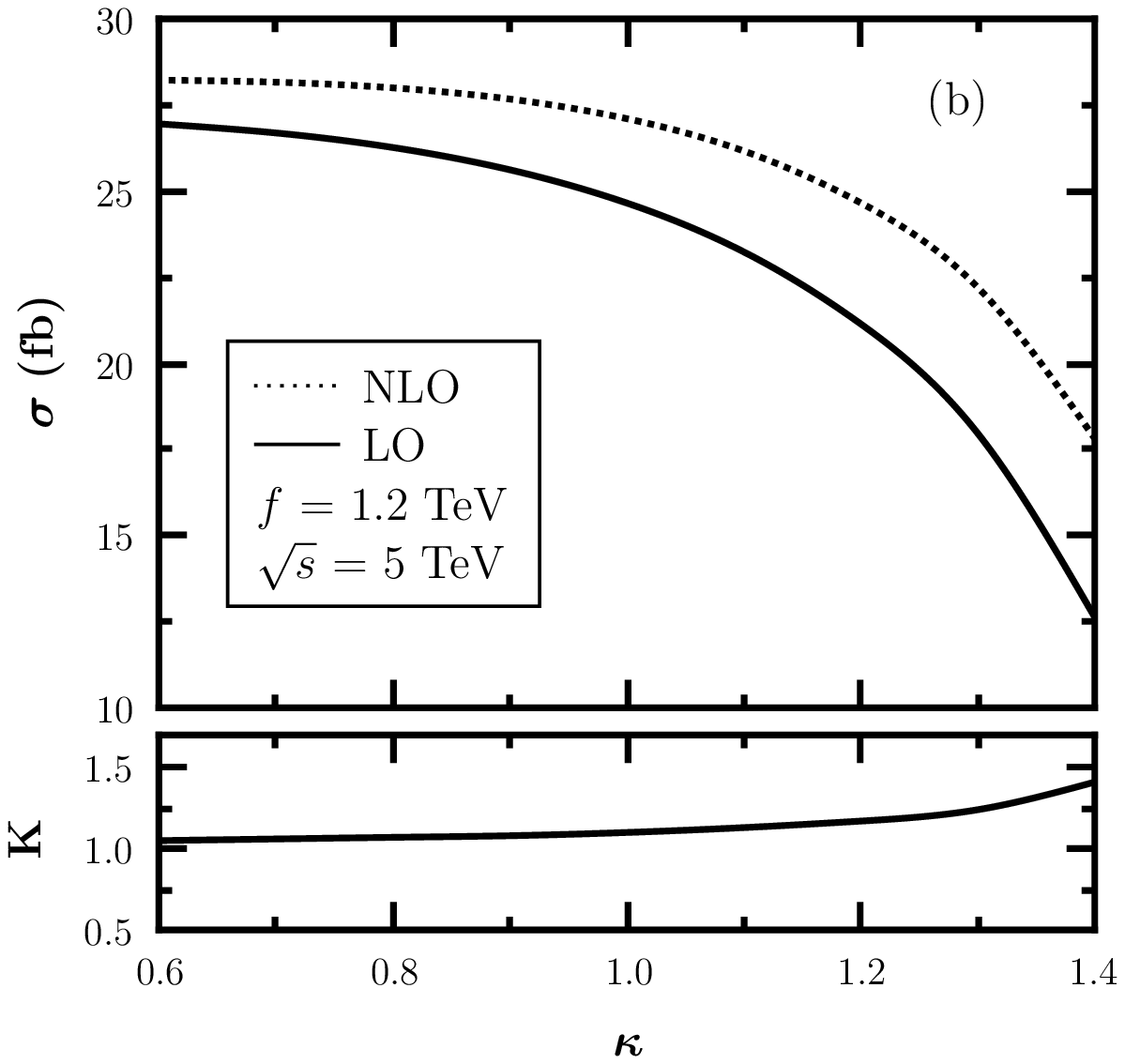}
\\~~  \\
\includegraphics[scale=0.64]{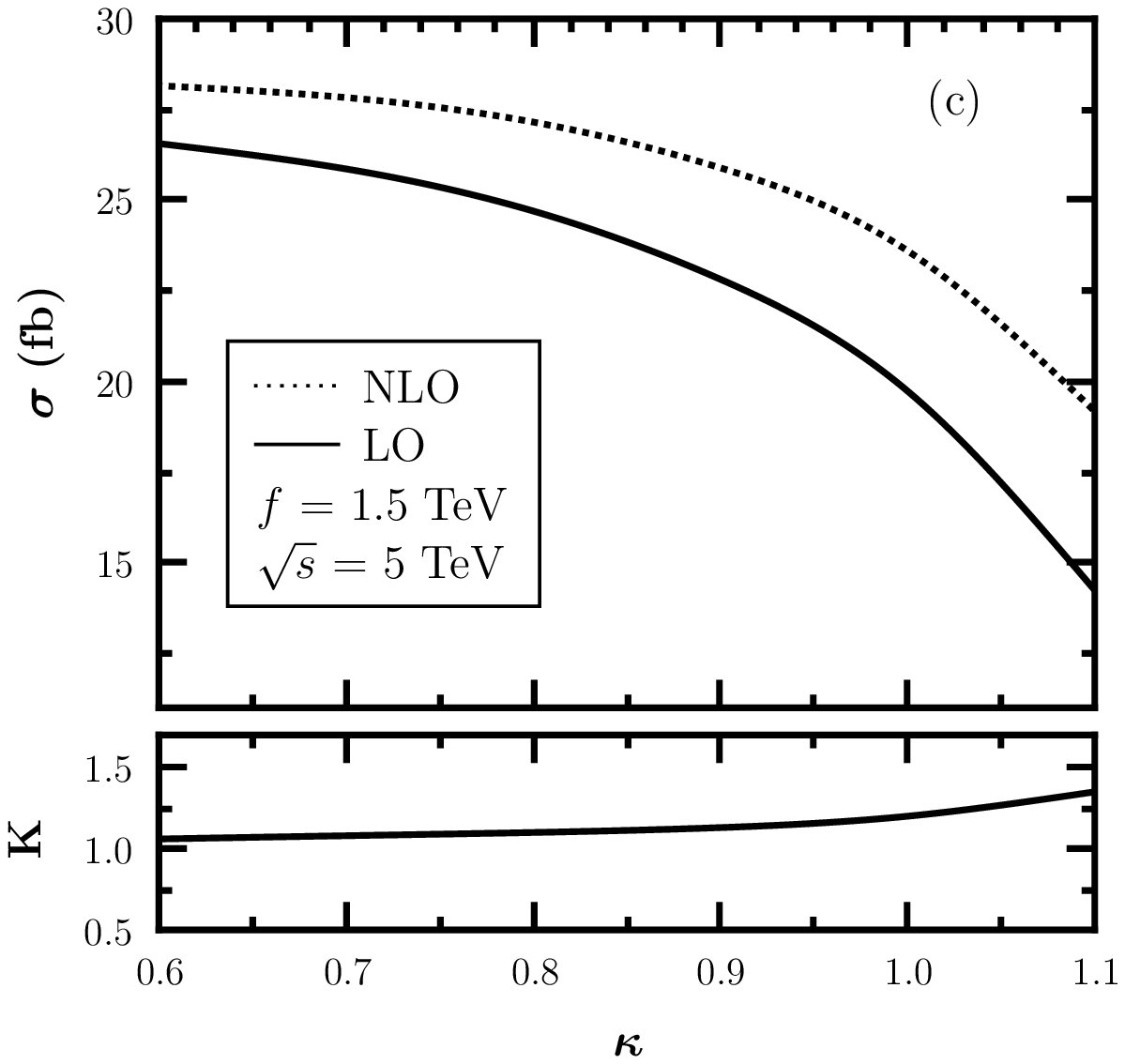}
\hspace{0in}%
\caption{\label{fig6-1} The LO, NLO QCD corrected cross sections and
the corresponding $K$-factors for \eeqq ($q_-\bar q_-=u_-\bar
u_-,~c_-\bar c_-,~d_-\bar d_-,~s_-\bar s_-$) as
functions of the $T$-odd mirror quark mass coefficient $\kappa$ at the
$\sqrt{s}=5~{\rm TeV}$ CLIC. (a) $f=1~{\rm TeV}$. (b) $f=1.2~{\rm TeV}$. (c)
$f=1.5~{\rm TeV}$.  }
\end{center}
\end{figure}

In Figs.\ref{fig6-1}(a), (b) and (c) we present the LO, NLO QCD
corrected integrated cross sections and the corresponding
$K$-factors as functions of the LHT $T$-odd quark mass
coefficient $\kappa$ at the $\sqrt{s}=5~{\rm TeV}$ CLIC by taking
$\mu_r=\mu_0$ and the global symmetry breaking scale as $f = 1,~1.2$
and $1.5~{\rm TeV}$, separately. We can see also from
Figs.\ref{fig6-1}(a,b,c) that the NLO QCD correction enhances the
integrated cross section, and the LO and NLO QCD corrected total
cross sections for \eeqq ($q_-\bar q_-=u_-\bar u_-,~c_-\bar
c_-,~d_-\bar d_-,~s_-\bar s_-$) at the $\sqrt{s}=5~{\rm TeV}$
CLIC decrease with the increment of $\kappa$, while the $K$-factor
increases with the increment of $\kappa$ due to the fact that the
threshold value approaches to the colliding energy when $\kappa$
goes up. We can read from the figures that for $f = 1~{\rm TeV}$ the
corresponding K-factor varies from $1.04$ to $1.15$ with
$\kappa$ going up from $0.6$ to $1.4$, for $f = 1.2~{\rm TeV}$ the $K$-factor
increases from $1.05$ to $1.41$ with $\kappa$ running from $0.6$ to
$1.4$, while for $f = 1.5~{\rm TeV}$ the $K$-factor goes up from $1.06$ to
$1.35$ with $\kappa$ increasing from $0.6$ to $1.1$.

\par
\subsection{Kinematic distributions of final decay products }
\par
\begin{figure}[htbp]
\begin{center}
\includegraphics[scale=0.74]{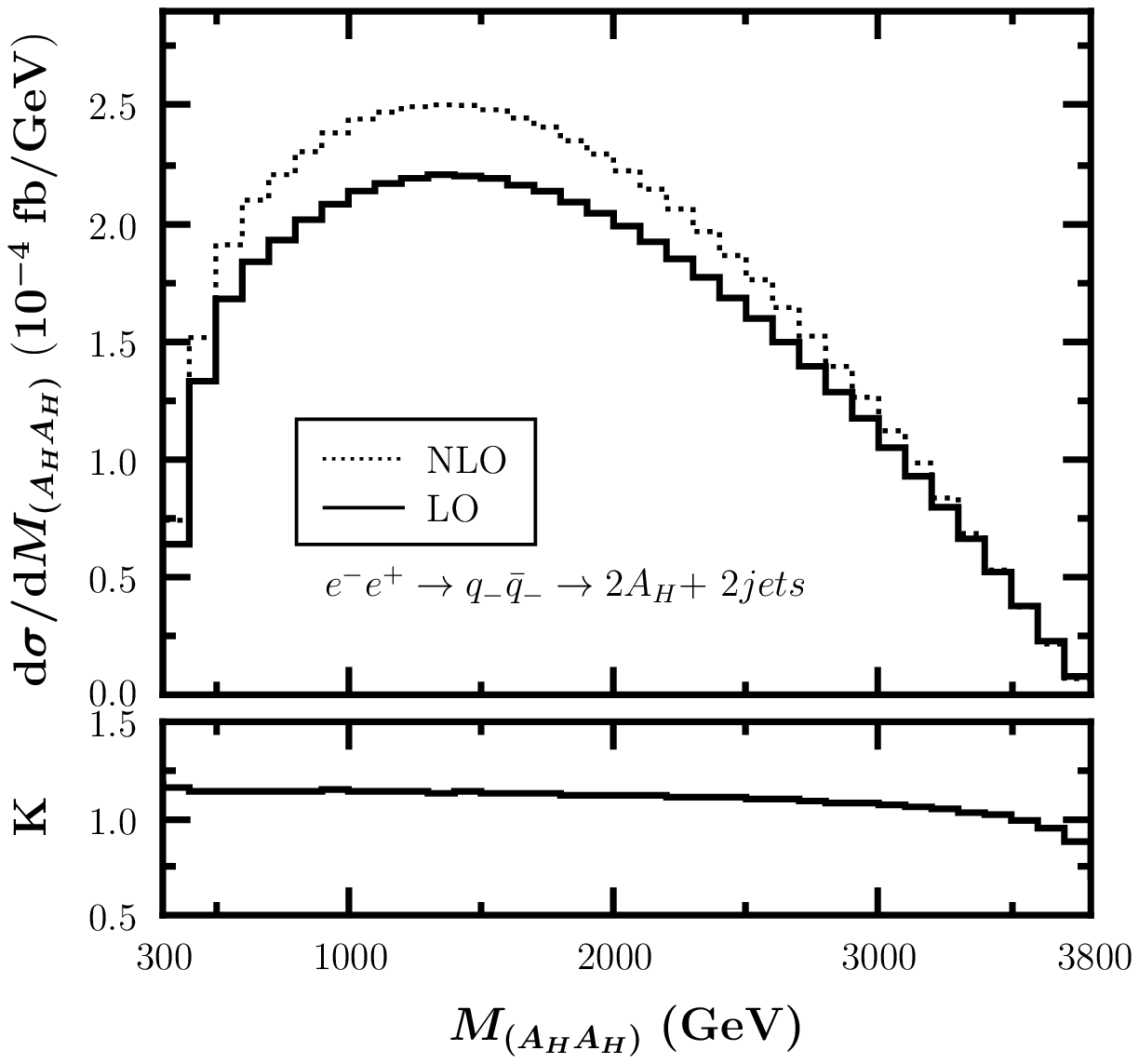}
\hspace{0in}%
\caption{\label{fig7} The LO, NLO QCD corrected distributions of the invariant mass of
$A_HA_H$ pair, and the corresponding $K$-factor for $e^+e^- \to q_-\bar q_- \to 2
A_H + 2 jets$ ($q_-=u_-,d_-,c_-,s_-$) at the
$\sqrt{s}=3~{\rm TeV}$ CLIC with $\kappa = 1$ and $f = 700~{\rm GeV}$. }
\end{center}
\end{figure}

We are interested in one of the final $T$-odd mirror quark $q_-$ decay
channels, i.e., $q_-(\bar{q}_-) \to A_H q(\bar{q})$. There we get
the $q_-(\bar{q}_-)$ decay products involving a $q(\bar q)$-jet and
missing energy of the lightest neutral stable particle $A_H$.
We assume the $q_-$ total decay width being the width summation of the
three main decay channels, i.e., $q_- \to V_H q$, ($V_H=Z_H,W_H,A_H$), and
the $q_- \to A_H q$ decay branch ratio up to the QCD NLO can be obtained
from Eqs.(\ref{Width-1}). We consider the $q_-\bar q_-$ pair
production process followed by the subsequential decay
$q_-(\bar{q}_-) \to A_H q(\bar{q})$ at the CLIC. We plot the LO and NLO
QCD corrected distributions of the final $A_HA_H$ pair invariant
mass for $e^+e^- \to q_-\bar q_- \to 2 A_H + 2 jets$
($q_-=u_-,d_-,c_-,s_-$) at the $\sqrt{s}=3~{\rm TeV}$ CLIC, and
the corresponding differential $K$-factor in Fig.\ref{fig7}. The
differential $K$-factor is defined as
\begin{eqnarray}
K(M_{(A_HA_H)})\equiv\frac{d\sigma_{NLO}/dM_{(A_HA_H)}}{d\sigma_{LO}/dM_{(A_HA_H)}}.
\end{eqnarray}
In this figure we take $f=700~{\rm GeV}$ and $\kappa=1$, and get $m_{A_H}
= 99.2~{\rm GeV}$. It demonstrates that the maximal differential cross
sections at both the LO and NLO are in the vicinity of
$M_{(A_HA_H)}\sim 1.3~{\rm TeV}$, and the differential $K$-factor varies
from $1.16$ to $0.88$ when $M_{(A_HA_H)}$ goes up from $300~{\rm GeV}$ to
$3800~{\rm GeV}$.

\par
\begin{figure}[htbp]
\begin{center}
\includegraphics[scale=0.6]{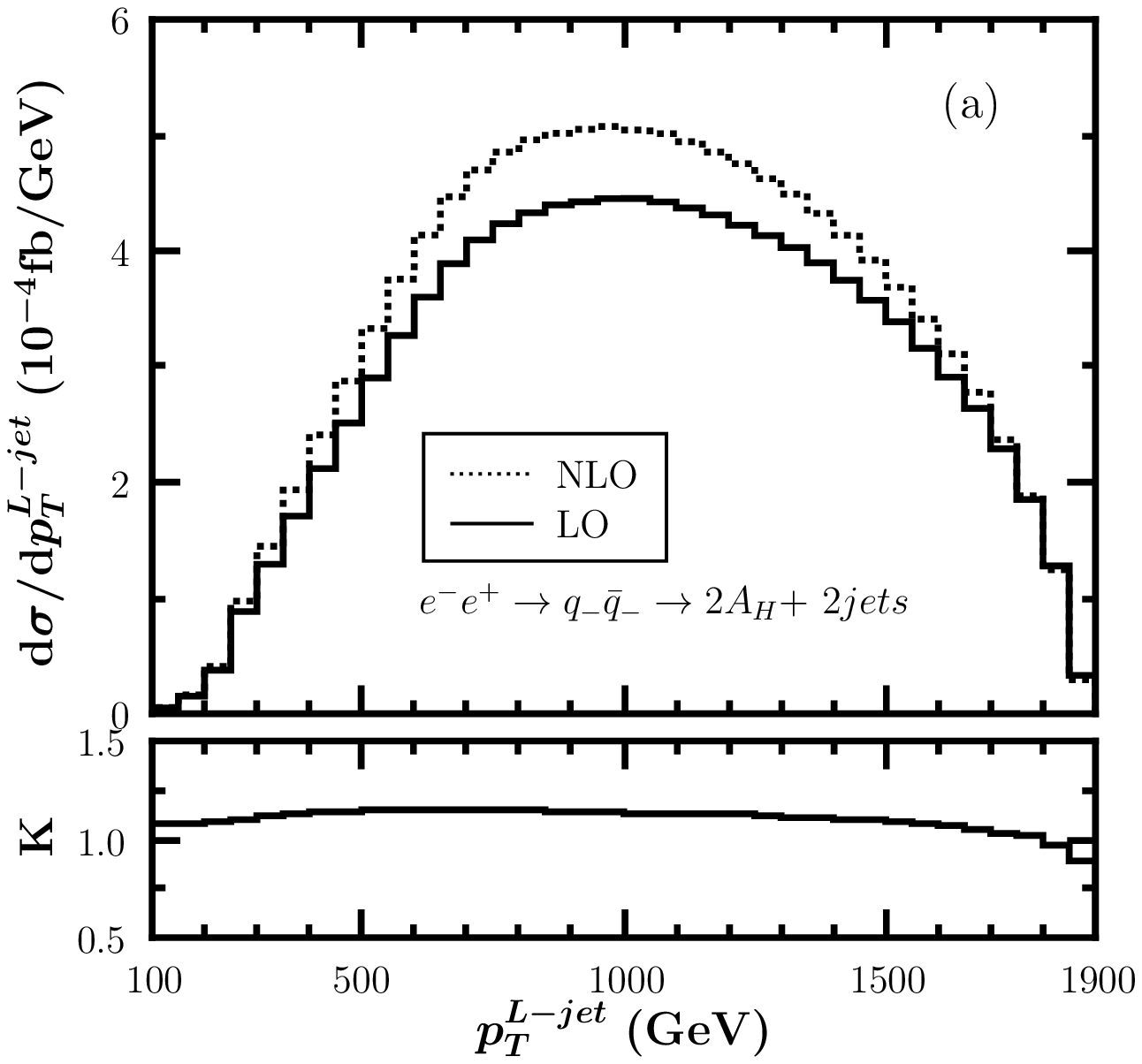}
\includegraphics[scale=0.6]{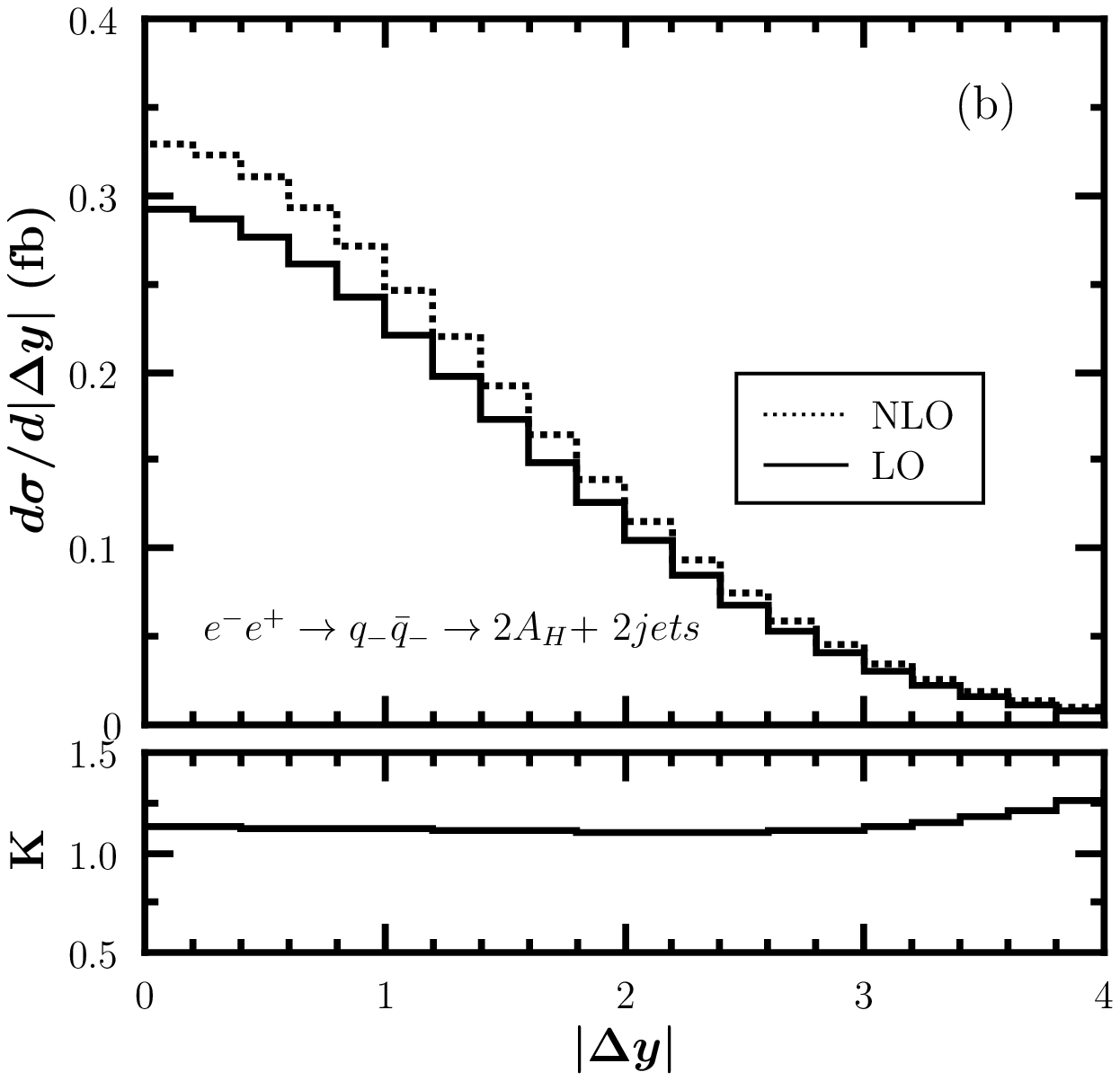}
\hspace{0in}%
\caption{\label{fig8} (a) The LO, NLO QCD corrected distributions
and the corresponding differential $K$-factor as functions of
the transverse momentum of the leading jet ($p_T^{L-jet}$) for
$e^+e^- \to q_-\bar q_- \to 2 A_H + 2 jets$
($q_-=u_-,~c_-,d_-,s_-$) at the $\sqrt{s} = 3~{\rm TeV}$ CLIC with
$\kappa = 1$ and $f = 700~{\rm GeV}$. (b) The LO, NLO QCD corrected
distributions and the corresponding $K$-factor for
$e^+e^- \to q_-\bar q_- \to 2 A_H + 2 jets$ ($q_-=u_-,~c_-,d_-,s_-$)
as functions of the rapidity separation of the final
leading jet and the next-to-leading jet $|\Delta
y|\equiv|y_{L-jet}-y_{NL-jet}|$ at the $\sqrt{s} = 3~{\rm TeV}$ CLIC with
$\kappa = 1$ and $f = 700~{\rm GeV}$. }
\end{center}
\end{figure}

For the LO and QCD NLO $e^+e^- \to q_-\bar q_- \to 2 A_H + 2 jets$
events, if the two or three jet (real gluon emission process) event
with a jet transverse energies satisfy the condition of
$E_{Tj_1}>E_{Tj_2}$ or $E_{Tj_1}>E_{Tj_2}>E_{Tj_3}$, we call $j_1$
as the leading jet and $j_2$ as the next-to-leading jet. In
Fig.\ref{fig8}(a) we depict the LO and NLO QCD corrected transverse
momentum distributions of the leading jet at the $\sqrt{s}=3~{\rm TeV}$
CLIC and the corresponding differential $K$-factor
($K(p_{T}^{L-jet})\equiv\frac{d\sigma_{NLO}/dp_{T}^{L-jet}}{d\sigma_{LO}/dp_{T}^{L-jet}}$).
There we take $f=700~{\rm GeV}$ and $\kappa=1$, and get $m_{A_H} =
99.2~{\rm GeV}$. The figure shows that the NLO QCD correction enhances the
LO distribution significantly. We see also that the peaks for the LO
and QCD NLO curves are located in the vicinity of $p_T^{L-jet}\sim
1~{\rm TeV}$, and the differential $K$-factor runs from $1.08$ to $0.88$
when $p_T^{L-jet}$ varies from $100~{\rm GeV}$ up to $1900~{\rm GeV}$.

\par
In Fig.\ref{fig8}(b) the LO and NLO QCD corrected distributions of
the rapidity separation of the final leading jet and the
next-to-leading jet ($|\Delta y|\equiv |y_{L-jet}-y_{NL-jet}|$), and
the corresponding differential $K$-factor ($K(|\Delta
y|)\equiv\frac{d\sigma_{NLO}/d|\Delta y|}{d\sigma_{LO}/d|\Delta
y|}$) are plotted. It shows that most of the $e^+e^- \to q_-\bar q_-
\to 2 A_H + 2 jets$ events are concentrated in the low $|\Delta y|$
region. We see also that the NLO QCD correction significantly
enhances the LO differential cross section, and the
differential $K$-factor varies between $1.13$ and $1.31$ with $|\Delta
y|$ in the range of $[0,4]$. All the distributions in
Fig.\ref{fig7} and Figs.\ref{fig8}(a,b) show that a simple
approximation of multiplying the LO distribution with the integrated
$K$-factor is not appropriate for precision study of the \eeqq
($q_-\bar q_-=u_-\bar u_-,~c_-\bar c_-,~d_-\bar d_-,~s_-\bar s_-$)
processes at the CLIC. It is necessary to calculate the complete NLO QCD
correction to get reliable distributions.

\vskip 5mm
\section{Summary}
\par
In this paper we present the precision calculations of the $T$-odd mirror quark pair
production including subsequential weak decays at the $e^+e^-$ CLIC up to the
QCD NLO in the littlest Higgs model with $T$-parity. The
future CLIC could provide an efficient facility to put the
precision measurements for this production process into practice with
a clean environment. The dependence of the NLO QCD effect to the
total cross section for the \eeqq ($q_-\bar q_-=u_-\bar u_-,~c_-\bar
c_-,~d_-\bar d_-,~s_-\bar s_-$) processes on colliding energy
$\sqrt{s}$ is investigated, and the LO and NLO QCD kinematic
distributions of final decay products are discussed. We find that
the NLO QCD correction always enhances the LO physical observables.
The $K$-factor is clearly related to the observable and
the phase space region, and increases obviously when the
production threshold approaches the colliding energy.
We see that the $K$-factor for the integrated cross
section varies in the ranges of  $1.10 \sim 1.39$ ($1.06 \sim 1.31$, $1.05 \sim 1.28$)
with $f \in [700,~ 1000]~{\rm GeV}$ ($f \in [700,~ 1300]~{\rm GeV}$,
$f \in [700,~ 1600]~{\rm GeV}$) at $\sqrt{s} = 3~{\rm TeV}$ ($4~{\rm TeV},~5~{\rm TeV}$) CLIC.
We conclude that NLO QCD correction has relevant impact on the $e^+e^- \to
q_-\bar q_- \to 2 A_H + 2 jets$ ($q_-=u_-,~c_-,d_-,s_-$) processes,
and should be included in any reliable analysis.

\vskip 5mm
\par
\noindent{\large\bf Acknowledgments:} This work was supported in
part by the National Natural Science Foundation of China (Grants.
No.11275190, No.11375008, No.11375171), and the Fundamental Research
Funds for the Central Universities (Grant. No.WK2030040044).

\end{document}